\documentstyle[aps,prb,epsfig,floats]{revtex}
\renewcommand{\narrowtext}{\begin{multicols}{2}\global\columnwidth20.5pc}
\renewcommand{\widetext}{\end{multicols}\global\columnwidth42.5pc}

\begin{document}

\draft

\twocolumn[\hsize\textwidth\columnwidth\hsize\csname@twocolumnfalse%
\endcsname

\title{The magnetic neutron scattering resonance of high-$T_{\rm c}$
       superconductors in external magnetic fields: an SO(5) study}

\author{
Niels Asger Mortensen$^{(1,3)}$, H. M. R{\o}nnow$^{(2,3)}$,
Henrik Bruus$^{(3)}$ and Per Hedeg{\aa}rd$^{(3)}$}

\address{
$^{(1)}$Mikroelektronik Centret, Technical University of Denmark, 
  {\O}rsteds Plads, Bld. 345 east, DK-2800 Kgs. Lyngby, Denmark\\
$^{(2)}$Condensed Matter Physics and Chemistry Department, Ris\o\
  National Laboratory, DK-4000 Roskilde, Denmark\\
$^{(3)}${\O}rsted Laboratory, Niels Bohr Institute for APG,
  Universitetsparken 5, DK-2100 Copenhagen {\O}, Denmark}

\date{June 15, 2000}
\maketitle

\begin{abstract}
The magnetic resonance at 41~meV observed in neutron scattering
studies of YBa$_2$Cu$_3$O$_7$ holds a key position in the
understanding of high-$T_{\rm c}$ superconductivity. Within the
SO(5) model for superconductivity and antiferromagnetism, we have
calculated the effect of an applied magnetic field on the neutron
scattering cross-section of the magnetic resonance. In the presence
of Abrikosov vortices, the neutron scattering cross-section shows clear
signatures of not only the fluctuations in the superconducting order
parameter $\psi$, but also the modulation of the phase of $\psi$ due
to vortices. In reciprocal space we find that i) the scattering 
amplitude is zero at $(\pi/a,\pi/a)$, ii) the resonance peak is split
into a ring with radius $\pi/d$ centered at $(\pi/a,\pi/a)$, $d$ being
the vortex lattice constant, and consequently, iii) the splitting
$\pi/d$ scales with the magnetic field as $\sqrt{B}$.
\end{abstract}

]


Soon after the discovery of high-$T_c$ superconductivity in the doped
cuprate compounds, its intimate relation to antiferromagnetism was
realized. A key discovery in the unraveling of this relationship was
the observation of the so called 41~meV magnetic 
resonance\cite{rossat-mignod91,mook93,fong95} later also
denoted the $\pi$ resonance. In inelastic neutron
scattering experiments on YBa$_2$Cu$_3$O$_7$ at temperatures below
$T_{\rm c}\sim 90\,{\rm K}$, Rossat-Mignod {\it et al.}
\cite{rossat-mignod91} found a sharp peak at $\hbar\omega\sim
41\,{\rm meV}$ and ${\bf q}=(\pi/a,\pi/a)$, $a$ being the lattice constant
of the square lattice in the copper-oxide planes.  Later its
antiferromagnetic origin was confirmed by Mook {\it et al.}
\cite{mook93} in a polarized neutron scattering experiment and
subsequently Fong {\it et al.\/}\cite{fong95} found that the magnetic
scattering appears only in the superconducting state.
Recently, Fong \emph{et al.}\cite{fong99b} have also observed the
$\pi$ resonance in Bi$_2$Sr$_2$CaCu$_2$O$_{8+\delta}$, which means
that it is a general feature of high-$T_c$ superconductors and not a
phenomenon restricted to YBa$_2$Cu$_3$O$_7$. This gives strong
experimental evidence for the $\pi$ resonance being related to
antiferromagnetic fluctuations within the superconducting state.
Conversely, it may be noted that angular-resolved photoemission
spectroscopy has shown how the single-particle gap within the
antiferromagnetic state inherits the $d$-wave modulation of the
superconducting state.\cite{ronning98,hanke99}

A number of different models have been proposed to explain the
$\pi$ resonance.
\cite{demler95,liu95,mazin95,zhang97a,abrikosov98,morr98,abanov99,brinckmann99}
In particular, Zhang was inspired by the existence of
antiferromagnetic fluctuations in the superconducting state to
suggest a unified SO(5) theory of antiferromagnetism and $d$-wave
superconductivity in the high-$T_{\rm c}$
superconductors.\cite{zhang97a} It is of great interest to extend
the different theoretical explanations to make predictions for the
behavior of the $\pi$ resonance \emph{e.g.} in an applied magnetic
field.  An experimental test of such predictions will put
important constraints on theoretical explanations of the $\pi$
resonance in particular and of high-$T_c$ superconductivity in
general.  In this paper we treat the $\pi$ resonance in the
presence of an applied magnetic field within the SO(5) model.

Zhang proposed that the cuprates at low temperatures can be understood
as a competition between $d$-wave superconductivity and
antiferromagnetism of a system which at higher temperatures possesses
SO(5) symmetry.\cite{zhang97a} The SO(5) symmetry group is the
minimal group that contains both the gauge group U(1) [$=$SO(2)]
which is broken in the superconducting state, and the spin rotation
group SO(3) which is broken in the antiferromagnetic state.
Furthermore, the SO(5) group also contains rotations of the
superspin between the antiferromagnetic sector and the superconducting
sector. The relevant order parameter is a real vector 
${\bf n}=(n_1,n_2,n_3,n_4,n_5)$ in a five dimensional superspin space
with a length which is fixed ($\left|{\bf n}\right|^2=1$) at low
temperatures. This order parameter is related to the complex
superconducting order parameter, $\psi$, and the antiferromagnetic
order parameter, $\bf m$, in each copper-oxide plane as follows: $\psi
= f e^{ i \phi}= n_1+i n_5$ and ${\bf m}=(n_2,n_3,n_4)$. Zhang argued
how in terms of the five dimensional superspin space one can construct
an effective Lagrangian ${\cal L}({\bf n})$ describing the low energy
physics of the $t$-$J$ limit of the Hubbard model. 

Two comments are appropriate here. 
  Firstly, we note that relaxing the constraint
$\left|{\bf n}\right|^2=1$ in the bulk superconducting state
will introduce high energy modes, but these can safely be ignored at
low temperatures. Moreover, they do not alter the topology of vortices
in the order parameter, which is our main concern. 
  Secondly, one may worry that results obtained from a pure SO(5)
model deviate substantially from those obtained from the recently
developed, physically more correct projected SO(5) theory
\cite{Zhang99}. However, the two models are only significantly
different close to half filling, and our study concerns AF-modes in 
the bulk superconductor in a weak magnetic field, a state which although 
endowed with the topology of vortices is far from half filling. 
For simplicity, we thus restrict the calculations in this paper 
to the original form of the SO(5) theory.

In the superconducting state the SO(5) symmetry is spontaneously
broken which leads to a ``high'' energy collective mode where the
approximate SO(5) symmetry allows for rotations of $\bf n$ between the
superconducting and the antiferromagnetic phases.  These
rotations have an energy cost $\hbar \omega_\pi$ corresponding to the
$\pi$ resonance and fluctuations in $\bf n$ will thus give rise to a
neutron scattering peak at $\hbar\omega_\pi$ which, through the
antiferromagnetic part of the superspin, is located at ${\bf q}={\bf
  Q}$, where ${\bf Q}=(\pi/a,\pi/a)$ is the antiferromagnetic ordering
vector. The uniform superconducting state ($f=1$) can be characterized
by a superspin ${\bf n}=(f \cos\phi,0,0,0,f \sin\phi)$, and the $\pi$
mode is a fluctuation $\delta{\bf n}(t) \propto (0,0,0,f
e^{i\omega_\pi t},0)$ around the static solution, where $\hat{\bf z}$
has been chosen as an arbitrary direction for $\delta{\bf m}$. In this
case with $f=1$ we have $\delta{\bf m}\propto 
e^{i\omega_\pi t}$, {\it i.e.\/} a sharp peak at $\omega=\omega_\pi$ and
${\bf q}={\bf Q}$. 

In the presence of an applied magnetic field, the superconductor
will be penetrated by flux quanta, each forming a vortex with a
flux $h/2e$ by which the complex superconducting order parameter
$\psi$ acquires a phase shift of $2\pi$ when moving around the
vortex.  In YBa$_2$Cu$_3$O$_7$ the vortices arrange themselves in
a triangular vortex lattice \cite{gammel87} with an area of the
hexagonal unit cell given by ${\cal A}=h/2eB$ and consequently a
lattice constant given by $d=3^{-1/4}\sqrt{h/eB}$. In the work by
Arovas {\it et al.},\cite{arovas97} Bruus {\it et
al.},\cite{bruus99} and Alama {\it et al.}\cite{alama99} the
problem of Abrikosov vortices was studied within the SO(5) model
of Zhang.\cite{zhang97a} In the center of a vortex core, the
superconducting part of the order parameter is forced to zero.
This leaves two possibilities: i) either the vortex core is in a
metallic normal state (as it is the case in conventional
superconductors) corresponding to a vanishing superspin or ii) the
superspin remains intact but is rotated from the superconducting
sector into the 
antiferromagnetic sector.\cite{arovas97} The prediction of the
possibility of antiferromagnetically ordered insulating vortex
cores is thus quite novel and allows for a direct experimental
test of the SO(5) theory. However, the antiferromagnetic ordering
of vortices is according to our knowledge still to be confirmed
experimentally. In this paper we report a different consequence of
the SO(5) theory in neutron scattering experiments; we consider
the $\pi$ mode in the presence of vortices and show that the peak
at ${\bf q}={\bf Q}$ splits into a ring with a radius $\pi/d$ centered
at ${\bf q}={\bf Q}$ where it has zero amplitude. Consequently
the splitting scales with magnetic field $B$ as $\pi/d\propto
\sqrt{B}$.

We start by considering just one vortex, then generalize the result to
a vortex lattice.  To make our calculations quantitative, we consider
YBa$_2$Cu$_3$O$_7$ for which $a= 3.8\,{\rm \AA}$, $\kappa \simeq 84$,
and $\xi\simeq 16\,{\rm \AA}$ for the lattice constant, the
Ginzburg--Landau parameter, and the coherence length, respectively.
The order parameter can be written in the form \cite{bruus99}

\begin{equation}
{\bf n}({\bf r})=
\big(\,f(r)\cos\phi_{\bf r},0,m(r),0,f(r)\sin\phi_{\bf r}\,\big),
\end{equation}
where $\phi_{\bf r}=\arg({\bf r})$. The isotropy of the
antiferromagnetic subspace allows us to choose $\bf m$ to lie in the
$y$-direction without loss of generality.  Static numerical solutions
for $f(r)$ and thereby also $m(r)$ in the presence of a vortex are
derived as described in Refs.  \onlinecite{arovas97,bruus99}.  Due to
the high value of $\kappa$ the absolute value $f$ of the
superconducting order parameter $\psi$ increases from zero at the
center of the vortex ($r=0$) to its bulk value ($f=1$) at a distance
of the order $\xi$ from the center.  The antiferromagnetic order
parameter follows from $f$ since $m=\sqrt{1-f^2}$.

For the $\pi$ mode in the presence of a vortex, 
Bruus {\it et al.} \cite{bruus99} found that the 
fluctuation of the superspin is
\begin{equation}
\delta{\bf n}({\bf r},t)=
\big(0,0,0,\delta\theta\,f(r)\cos\phi_{\bf r}\, e^{i\omega_{\pi} t},0\big),
\end{equation}
where the small angle $\delta\theta$ by which $\bf n$ rotates into the
antiferromagnetic sector is undetermined. Since the excitation depends
on $f$ and not on $m$ it is a de-localized excitation with zero
amplitude at the center of the vortices and in terms of energy it
actually corresponds to an energy at the bottom edge of the continuum
of an effective potential associated to the vortices.\cite{bruus99}

For an isotropic spin space, the magnetic scattering cross-section for
neutrons is proportional to the dynamic structure factor, which is the
Fourier transform of the spin-spin correlation function (see {\it
  e.g.} Ref. \onlinecite{squires78}),
\begin{equation}
{\cal S}({\bf q},\omega)=
\int_{-\infty}^\infty {\rm d}t\, e^{i\omega t}
\sum_{\bf RR'}e^{-i {\bf q}\cdot({\bf R}-{\bf R'})}
\left<\hat{\bf S}_{\bf R}(t)\cdot\hat{\bf S}_{{\bf R'}}(0) \right>.
\end{equation}

To make a connection to the SO(5) calculations we make the
semiclassical approximation $\big< \hat{\bf S}_{\bf R}(t)\cdot\hat{\bf
  S}_{{\bf R}'}(0) \big> \approx\big<\hat{\bf S}_{\bf R}(t)\big>
\cdot\big<\hat{\bf S}_{{\bf R}'}(0) \big>$ so that

\begin{eqnarray}
{\cal S}({\bf q},\omega)&\approx&
\int_{-\infty}^\infty {\rm d}t\, e^{i\omega t}
\sum_{{\bf R},{\bf R}'} e^{-i \left({\bf q}+{\bf
Q}\right)\cdot({\bf R}-{\bf R}')}\nonumber\\ &\quad&\times
{\bf m}({\bf R},t)\cdot {\bf m}({\bf R}',0),
\end{eqnarray}
where ${\bf m}({\bf R},t)=e^{i{\bf Q}\cdot{\bf R}}{\bf S}_{\bf R}(t)$
is the antiferromagnetic order parameter which enters the superspin
$\bf n$.

With a superspin given by ${\bf n}({\bf r},t)={\bf n}({\bf
  r})+\delta{\bf n}({\bf r},t)$ the dynamical structure factor has two
components --- an elastic and an inelastic. The elastic component
\begin{equation}
{\cal S}_{\rm el}({\bf q},\omega)=
\left|
\sum_{\bf R} e^{-i({\bf q}+{\bf Q})\cdot{\bf R} }m(R)
\right|^22\pi\delta(\omega),
\end{equation}
is located at ${\bf q}={\bf Q}$ and has a width $\sim \pi/\xi$.
In elastic neutron scattering experiments the observation of this peak
would directly prove the antiferromagnetical ordering in vortex cores.

The inelastic contribution is
\begin{eqnarray}
{\cal S}_{\rm in}({\bf q},\omega)&=&
\left(\delta\theta\right)^2
\left| \sum_{\bf R} e^{-i({\bf q}+{\bf Q}){\bf R} }
\,f(R)\cos\phi_{\bf R}
\right|^2 \nonumber\\
&\quad& \times 2\pi\delta(\omega-\omega_\pi).
\label{structurefactor}
\end{eqnarray}
For ${\bf q}={\bf Q}$ the phase factor $e^{-i({\bf q}+{\bf Q}){\bf
R}}$ vanishes, and the cosine factor makes the different terms in
the summation cancel pairwise so that ${\cal S}_{\rm in}({\bf
  Q},\omega_\pi)=0$.  The presence of a single vortex moves the
intensity away from ${\bf q}={\bf Q}$ and a ring-shaped peak
with radius $\delta q \sim\pi/L$ centered at ${\bf q}={\bf Q}$
is formed, $L\sim\sqrt{A}$ being the size of the sample.  
In the semiclassical approximation the zero amplitude at 
${\bf q}={\bf Q}$ is a topological feature, which is independent of
the detailed radial form $f(r)$ of the vortex. This robustness relies
on the identification of the $\pi$ mode as being proportional to the
superconducting order-parameter (including its phase). Quantum
fluctuations may add some amplitude at ${\bf q}={\bf Q}$, but such an
analysis beyond leading order is outside the scope of this work. 

It is interesting to see how this result compares to predictions based
on the BCS theory. The neutron scattering cross-section is given by
the spin susceptibility, which for a homogeneous (vortex free)
superconductor has been calculated via the BCS-Lindhard
function.\cite{liu95,mazin95} Here we briefly consider how
the BCS coherence factor $[u_k v_{k+q} - v_k u_{k+q}]^2$ appearing in
the Lindhard function\cite{schrieffer} is modified by the presence of
vortices. In a semiclassical approximation\cite{volovik} the spatial
variation of the superconducting phase $\phi({\bf r})$ leads to a
coherence factor of
the form $[u_k({\bf r}_1)     e^{i\phi({\bf r}_1)/2}
           v_{k+q}({\bf r}_2) e^{i\phi({\bf r}_2)/2} -
           v_k({\bf r}_1)     e^{i\phi({\bf r}_1)/2}
           u_{k+q}({\bf r}_2) e^{i\phi({\bf r}_2)/2}]^2$. Therefore in
contrast to Eq.~(\ref{structurefactor}) the superconducting phase does
not separate in the two spatial positions, and consequently the
spatial average in general is not zero at ${\bf q}={\bf Q}$.
It thus appears that the above mentioned ring-shaped peak in the
dynamic structure factor is special for the SO(5) model.

\begin{figure}[t]
\begin{center}
\epsfig{file=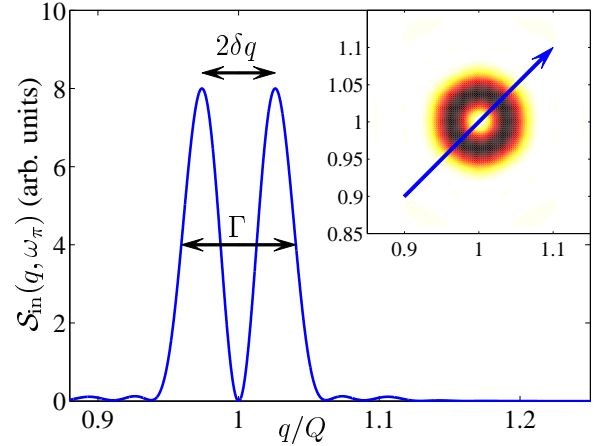, width=0.9\columnwidth,clip}
\end{center}
\caption{Plot of the dynamic structure factor at $\omega=\omega_\pi$ as a
  function of $q$ along the $(\pi,\pi)$-direction for $B=10\,{\rm T}$.
  The inset shows the almost isotropic response in the $q$-plane with
  the arrow indicating the $(\pi,\pi)$-direction.}
\label{FIG1}
\end{figure}

We now generalize the single-vortex SO(5)-result to the case of a
vortex lattice.  For non-overlapping vortices we construct the full
superconducting order parameter by

\begin{equation}
\tilde{\psi}({\bf r})=
\tilde{f}({\bf r})e^{i\tilde{\phi}({\bf r})}=
{\textstyle \prod_j} \psi({\bf r}-{\bf r}_j),
\end{equation}
where the ${\bf r}_j$ denote the positions of the vortices. The
function $\tilde{f}({\bf r})=
\prod_j f({\bf r}-{\bf r}_j)$ is $1$ except
for close to the vortices where it dips to zero.  Also the phase
$\tilde{\phi}({\bf r})=\sum_j \arg({\bf r}-{\bf r}_j)$ has by
construction the periodicity of the vortex lattice (modulo $2\pi$) and
the contour integral $\oint_C 
{\rm d}{\bf l}\!\cdot\!\mbox{\boldmath $\nabla$} 
\tilde{\phi}({\bf r})$ equals $2\pi n$ where
$n$ is the number of vortices enclosed by the contour $C$. In the
limit of non-overlapping vortices we can capture the main physics by
considering the single vortex solution within a unit cell of the
vortex lattice. We comment on the inclusion of the entire vortex
lattice further on, but for now we restrict the summation in
Eq.~(\ref{structurefactor}) to lattice sites $\bf R$ inside the vortex
lattice unit cell. In Fig.~\ref{FIG1} we show the result for a
magnetic field $B=10\,{\rm T}$. As seen, the presence of vortices
moves the intensity away from ${\bf q}={\bf Q}$ and a ring
shaped peak with radius $\delta q$ centered at ${\bf q}={\bf Q}$
is formed.  We note that the only relevant length scale available is
the vortex lattice constant $d$ and consequently we expect that
$\delta q=\pi/d$. Since $d=3^{-1/4}\sqrt{h/eB}$ we consequently expect
that $\delta q=3^{1/4}\pi\sqrt{eB/h}\simeq 0.008\times
(\pi/a)\sqrt{B/[{\rm T}]}$. Had we included all the vortex lattice
unit cells in our analysis, the structure factor of the hexagonal
vortex lattice would have led to a breaking of the ring in
Fig.~\ref{FIG1} into six sub-peaks sitting on top of the ring. In a
real experiment these sub-peaks could easily be smeared back into a
ring-shaped scattering peak if either the vortex lattice were slightly
imperfect or if the resolution of the spectrometer were too low. To
describe the main effect of the SO(5) theory we therefore continue to
use the single unit cell approximation.

\begin{figure}[t]
\begin{center}
\epsfig{file=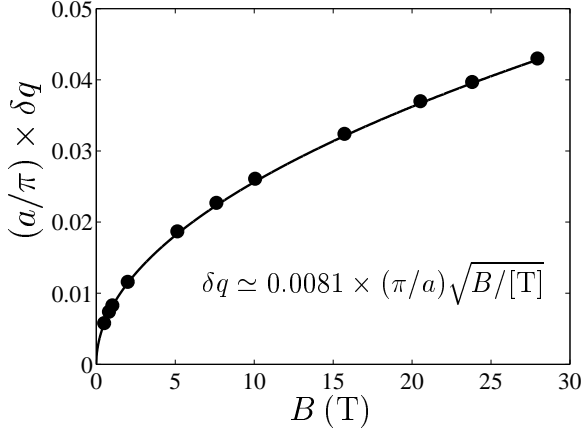, width=0.9\columnwidth}
\end{center}
\caption{Plot of the peak splitting $\delta q$ as a function of the
  magnetic field $B$. The calculated splitting ($\bullet$) has the
  expected $B^{1/2}$ behavior and the numerical pre-factor
  corresponds to a splitting $\delta q =\pi/d$ where $d$ is the vortex
  distance.}
\label{FIG2}
\end{figure}

In Fig.~\ref{FIG2} we show the splitting
as a function of the magnetic field and indeed we find the expected
scaling with a pre-factor confirming that the splitting is given by
$\delta q=\pi/d$. The full width half maximum of the ring is given by
$\Gamma \simeq 3.1\times \delta q= 3.1\times \pi/d$.

\begin{figure}[t]
\begin{center}
\epsfig{file=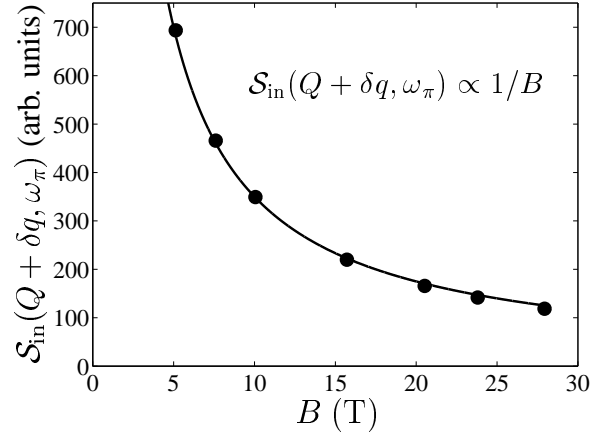, width=0.9\columnwidth}
\end{center}
\caption{The peak height plotted versus the magnetic
  field $B$. The calculations ($\bullet$) almost fit a $1/B$ dependence.}
\label{FIG3}
\end{figure}

In Fig.~\ref{FIG3} we show the amplitude of the ring as a function of
magnetic field. The amplitude approximately decreases as $1/B$ with the
magnetic field, but with a small deviation.  This deviation makes the
$\bf q$-integrated intensity, which is proportional to the amplitude
times $(\delta q)^2$, decrease as $I(B)/I(0)\simeq 1-0.004 \times
B/[{\rm T}]$ which reflects that the area occupied by vortices
increases linearly with $B$ and consequently the superconducting
region decreases linearly with $B$. In fact, the reduction is given by
${\cal A}^{-1} 2\pi\int r{\rm d}r\,m^2(r) \simeq 0.004 \times B/[{\rm
  T}]$, where the integral gives the effective area of the vortex.
The reduction in integrated intensity should be relatively easy to
observe experimentally, but is not a unique feature of the SO(5)
model. Thus, while it will aid to prove that the $\pi$ resonance only
resides in the superconducting phase, it will not clearly distinguish
between different theories.

In order to discuss the experimental possibilities for testing our
predictions, we note that the original observation of the zero-field
$\pi$ resonance was an experimental achievement and hence that
the experiment proposed here constitutes a great challenge. 
However, since the first observation of
the $\pi$ resonance in 1991, the field of neutron scattering has
developed considerably. To observe the ring-like shape (see inset of
Fig.~\ref{FIG1}) of the excitation would require a resolution better
than $\pi/d$ along two directions in reciprocal space, which seems
unachievable with current spectrometers. However, the overall width of
the ring can in fact be measured with good resolution along just one
direction in the reciprocal plane.
Scans along this direction (as in Fig.~\ref{FIG1})
could then reveal a broadening of $\sim3.1\times\pi/d$. With a
sufficiently optimized spectrometer we believe this to be possible,
and the reward is a stringent test of a quantitative prediction of the
SO(5) theory. We note that Bourges {\it et al.} \cite{bourges97b} have
investigated the $\pi$ resonance in a magnetic field of $B=11.5\,{\rm
  T}$ and report a broadening in energy, but do not report data on the
${\bf q}$-shape.

In conclusion we have found that within the SO(5) model, the $\pi$
resonance splits into a ring centered at ${\bf q}=(\pi/a,\pi/a)$ in
the presence of a magnetic field.  The ring has the radius $\pi/d$ and
full width half maximum of about $3.1\times \pi/d$, where $d$ is the
vortex lattice constant. Consequently the splitting is found to scale
with the magnetic field as $B^{1/2}$. We emphasize that the
  amplitude of the $\pi$ resonance is zero at ${\bf q}=(\pi/a,\pi/a)$
  in the presence of a magnetic field.

We acknowledge useful discussions with J. Jensen, N. H. Andersen,
A.-P. Jauho and D. F. McMorrow. H.M.R. is supported by the Danish
Research Academy and H.B. by the Danish Natural Science Research
Council through Ole R\o{}mer Grant No.  9600548.

\end{document}